# Numerical Investigation of Coaxial GCH$_4$/LOx Combustion at Supercritical Pressures

Sindhuja Priyadarshini[1], Malay K Das[2], Ashoke De[1*], Rupesh Sinha[3]

[1]Department of Aerospace Engineering, Indian Institute of Technology, Kanpur, India - 208016
[2]Department of Mechanical Engineering, Indian Institute of Technology, Kanpur, India -208016
[3]Liquid Propulsion Systems Center, ISRO, Valiamala, Thiruvananthapuram, India -695547

*Corresponding Author: Tel.: +91-512-2597863 Fax: +91-512-2597561

E-mail address: ashoke@iitk.ac.in

**ABSTRACT**

This article aims to numerically investigate the combustion phenomenon of coaxial gaseous CH$_4$/LOx at supercritical pressures. The choice of turbulence model, real gas model, and chemical kinetics model are the critical parameters in numerical simulations of cryogenic combustion at high pressure. At this supercritical operating pressure, the ideal gas law does not remain valid for such cases. Therefore, we have systematically carried out a comparative study to analyze the importance of real gas models, turbulence parameters, and chemical kinetics at such conditions. The comparison of real gas models with the NIST database reveals better conformity of SRK (Soave Redlich Kwong – Equation of State (EoS)) model predictions with the database. Further, the computed results indicate that the Standard k-ε turbulence model with modified constant ($C_{\varepsilon 1}$ = 1.4) captures the better flame shape and temperature peak position compared to other RANS based turbulence models while invoking the non-premixed steady β-PDF flamelet model for simulating the combustion process. Furthermore, a comparative study comparing two different chemical kinetics models indicates that the reduced Jones-Lindstedt mechanism (JL-R) can accurately predict the flame characteristics with the least computational cost. Finally, we have studied the effect of chamber pressure and LOx inlet temperature on the flame characteristics. The flame characteristics exhibit a strong sensitivity towards the chamber pressure due to the weakening of the pseudo-boiling effect with an increase in pressure. As a consequence of lower turbulent rates of energy and mass transfer through the transcritical mixing layer, the flame spreading becomes narrower at elevated pressure and temperature, thereby yielding an increased flame length at transcritical conditions.

**Introduction**

In the last few years, there has been an increasing interest in numerical modeling of combustion phenomenon in cryogenic engines due to its complex nature. The combination of liquid hydrogen (fuel) with liquid oxygen (oxidizer) has been widely utilized as rocket fuel and oxidizer for various liquid propulsion systems. Liquid hydrogen (fuel) has multiple advantages like non-toxicity, clean combustion, and the highest specific impulse. But, the low density of $H_2$ (liq.) leads to a large vehicle, a larger tank volume, and higher aerodynamic drag. Moreover, high cost and handling difficulties of $H_2$ (liq.) have prohibited the widespread use of $H_2$ (liq.)-LOx combination in liquid rocket engines (LRE's) (Sutton, 2006). Lately, it has been widely recognized that hydrocarbons are the most effective alternate propellants due to their high-density characteristics resulting in minimization of the propellant tank size and overall operational cost. The lowest hydrocarbon, liquid methane, has inherent properties like higher specific impulse and better cooling capabilities. The various advantages of liquid methane over other higher hydrocarbons have made it the most competitive fuel in combination with the liquid oxygen. Due to its soft cryogenic like characteristics, the $GCH_4$/LOx combination can easily be operated at a cryogenic arrangement.

There are studies related to combustion characteristics of LREs; their dynamics and behavior are available in the literature. Extensive experimental investigations have been performed to study the cold flows (injection of liquid nitrogen, $LN_2$, with gaseous nitrogen, $GN_2$) in order to gain an in-depth understanding of the flow behavior and to discover its influence on the flow field structures (Chehroudi et al., 2002; Chehroudi et al., 2002; Mayer et al., 2003; Oefelein, 2006; Oschwald et al., 2006). Ge et al. (2019), for instance, performed a high-pressure combustion experiment to study the flame characteristics of methane/oxygen laminar co-flow diffusion flame. They stated that three combustion states are present in the flow field namely the steady combustion state, transition combustion state, and the pulsation combustion state.

Two major research groups in Europe experimentally investigate the turbulent combustion of $H_2$/LOx and $CH_4$/LOx: (1) ONERA that developed the Mascotte cryogenic test bench (Zurbach et al., 2002), (2) DLR. Lampoldshausen, Germany using the M3 burner (Yang et al., 2007). ONERA developed the Mascotte cryogenic test bench, which is regarded as one of the most common configurations to study combustion phenomena at supercritical pressure. Remarkable experimental investigations were carried out by Candel et al. (2006) and Singla et al. (2005) on the ONERA facility to study the flame characteristics at high pressure. They reported that the flame structure is mainly affected by the turbulent mass transfer processes from the central core due to the velocity difference between two streams. In a subsequent study, they (Singla et al., 2007) further compared the flame structures formed by different fuel combinations such as $CH_4$/LOx and $H_2$/LOx. They found that in the case of $H_2$/LOx, the OH layer has a thickness which is of the order of the lip thickness, and the anchor point is located close to the lip. Whereas in the case of $CH_4$/Lox, they observed that the OH layer thickness exceeds the lip thickness and the anchor point is located downstream from the lip. Owing to the experimental data that are available from the Mascotte cryogenic test bench, several research groups compared their numerical results with the same database to shed light on the modeling aspect of such combustion phenomena (Cutrone et al., 2006; Cutrone et al., 2010; De Giorgi et al., 2010; Ficarella et al., 2009).

In numerical modeling of combustion dynamics in LRE's, besides the classical difficulties in dealing with turbulent reacting flows, new phenomena and problems arise due to the departures of thermodynamic properties from their ideal-gas limit, namely the vanishing of surface tension and the enthalpy of vaporization for supercritical fluids. At such conditions, the reactant properties show liquid-like densities, gas-like diffusivities, and pressure-dependent solubility. The critical pressure of oxygen and methane are 5.043 MPa and 1.313 MPa, respectively. The real gas model has a strong impact on the flame structure at high-pressure conditions. Thus, the choice of the real gas model is one of the critical parameters that assist in better prediction of the combustion phenomena at supercritical pressure.

Kim et al. (2011) numerically investigated the flame features of nitrogen jets at near critical and supercritical pressures. Their principal focus was to choose a suitable cubic EoS model which is able to predict the essential features of the cryogenic liquid nitrogen jets. Their predictions showed that the predictions of the Peng Robinson (PR) model are in better conformity with the measured density profiles, compared to the Soave Redlich Kwong (SRK) model. Furthermore, they also investigated $H_2$/LOx coaxial jet flame at supercritical pressure using Soave Redlich-Kwong (SRK) EoS to capture the non-ideal thermodynamic behavior (Kim et al., 2011). In their further study, Kim et al. (2013) numerically investigated $CH_4$/LOx flame structures to show case the effect of three different EoS such as Soave Redlich Kwong (SRK) model, Peng Robinson (PR) model and ideal gas. They found that SRK EoS correctly predicts the thermodynamic variation at supercritical pressure. In a recent study, Mueller et al. (2015) developed a numerical method to perform large-eddy simulations (LES) of non-premixed LOx/CH$_4$ combustion at supercritical pressures by using Peng-Robinson EoS model for thermodynamic non-idealities and a steady laminar flamelet approach for turbulent combustion. Later, the same authors proposed a new real-gas flamelet model with increased numerical performance (Zips et al., 2018).

Despite these studies, there are recent developments that are made in real gas models. The volume translation methods for cubic EoS are recently implemented in the field of real gas models. Based on the general form of a cubic EoS, a mathematical framework for applying volume translations is provided, allowing a unified and thermodynamically consistent formulation for better prediction of the flame structure at supercritical pressure (Matheis et al., 2016). This methodology is applied to two different volume translation methods which were proposed by Abudour et al. (2012) and Baled et al. (2012). As emphasized, the main challenges of LOx/GCH$_4$ combustion at the supercritical pressure are the accurate representation of thermodynamic non-idealities. Hence, there is a clear need to establish a unique numerical framework within which the effects of all known models and design attributes can be studied and assessed using advanced modeling and simulation techniques. In the present work, we have initially carried out a comparative study using five different EoS viz. Soave Redlich Kwong (SRK) model, Redlich Kwong (RK) model, the Peng-Robinson (PR) model, ideal gas, SRK translated EoS, and PR translated EoS. The predictions of different EoS are compared against NIST data, and the best suited one is selected for detailed analysis to simulate the combustion phenomena at supercritical pressure.

In addition, the choice of the combustion model is another essential parameter to accurately predict the combustion phenomena at supercritical pressure. De Giorgi et al., (2009, 2011a, 2011b, 2012) compared the different combustion models for modeling of combustion phenomena in GCH$_4$/LOx LRE and concluded that the β-PDF based flamelet combustion model predicted the flame shape similar to the experimental observations at a reasonable computational cost and time. Moreover, the choice of chemical kinetics plays an important role in the accurate representation of combustion characteristics at supercritical pressure. Previous literature suggests that not all chemical mechanisms are suitable for oxygen-methane combustion (De Giorgi et al., 2012) at supercritical pressures as the real combustion process involves a large number of species and reactions. Thus, in the current study, we have also assessed the effect of different kinetic mechanisms, and the suitable one is used for further simulations. A brief description of two chemical kinetics models based on the number of species and reactions have been summarized in Table 1.

Therefore, the broad objective of the current study is to compare (i) different real gas models (ii) different turbulence models (iii) two different chemical kinetics model to find a better model combination in all three different regions which is able to describe the combustion phenomena accurately at G2 test condition. Furthermore, an attempt is made to investigate the effect of the elevated pressure and LOx inlet temperature on the turbulent flame structure of methane and liquid oxygen by incorporating a more suitable oxy-methane chemical kinetics mechanism at elevated pressure. The numerical results are validated against the experimental study of Test case RCM-3 Mascotte Single Injector, developed at ONERA. This injector is recently adapted to study the $GCH_4$/LOx combustion version-4 (V04) (Singla et al., 2007), and condition G2 (Kim et al., 2013) has been chosen, as summarized in the following sub-section (Table 4).

**Mathematical formulation**

*Governing Equation*

We solve Favre averaged governing equations of mass, momentum, energy, and turbulence quantities which can be generically recast as:

$$\frac{D(\rho \check{\phi})}{Dt} = \nabla^2(\rho D \check{\phi}) + \langle S_\phi \rangle \tag{1}$$

Here $\rho$ is the density, and $D$ is the coefficient of scalar. The Favre averaged scalar in the turbulent flow field is given by $\widetilde{\phi}$ and $\langle S_\phi \rangle$ represents the mean scalar term of the scalar. We have invoked various RANS based turbulence models such as SKE (standard $k$-$\varepsilon$), RKE (Realizable $k$-$\varepsilon$), SST $k$-$\omega$ and RSM (Reynolds Stress model) for comparison (Menter, 1994; Saqr & Wahid, 2011; Wilcox, 1998), while the effect of kinetics at elevated pressure is analysed using three kinetics mechanism as tabulated below. Notably, both JL-R and SKEL mechanisms are valid for high pressure oxy-methane combustion, as mentioned in the literature (Frassoldati et al., 2009; Yang & Pope, 1998).

*Turbulence-Chemistry Interaction*

The local flame structure of the turbulent flames can be described as an ensemble of laminar and one-dimensional local structures. With unitary Lewis number, the steady equations can be expressed with respect to mixture fraction $f$ and scalar dissipation rate $\chi$ as follows:

$$0 = \frac{\chi}{2}\frac{\partial^2 Y_i}{\partial f^2} + \frac{\dot{\omega}_i}{\rho} \tag{2}$$

$$0 = \frac{\chi}{2}\frac{1}{C_p}\frac{\partial^2 h}{\partial f^2} - \frac{\chi}{2}\frac{1}{C_p}\sum_{k=1}^{N} h_k \frac{\partial^2 Y_k}{\partial f^2} - \frac{1}{\rho C_p}\sum_{k=1}^{N} h_k \dot{\omega}_k \tag{3}$$

The connection between the physical space and mixture fraction space is achieved through the scalar dissipation rate, which quantifies the deviation from equilibrium and is defined as:

$$\chi = 2D \left|\frac{\partial f}{\partial x_j}\right|^2 \tag{4}$$

The scalar dissipation rate varies along the flamelet and is modeled as Eq. (4) as depicted below:

$$\chi(f) = \chi_{st} exp\left[2\big(erfc^{-1}(2f_{st})\big)^2 - 2\big(erfc^{-1}(2f)\big)^2\right] \tag{5}$$

Where $erfc^{-1}$ stands for the inverse of the complementary error function. The turbulent flame brush is represented as the ensemble of diffusion flamelets, where the Favre averaged temperature, and species mass fraction of a turbulent flame can be determined as:

$$\widetilde{\emptyset} = \iint \emptyset(f,\chi_{st})\, p(f,\chi_{st})\, df\, d\chi_{st} \qquad (6)$$

Where a presumed β-PDF, used to define the probability of the mixture fraction. The temperature and mean density have an extra dimension of mean enthalpy $\widetilde{H}$ to consider the non-adiabatic steady diffusion flamelets. The species mass fraction is assumed to have an eligible effect by the heat loss or gain by the system. The evolution of mixture fraction in the physical space is represented by the transport equation of *f* and *f* " is given as:

$$\frac{\partial}{\partial t}(\rho \tilde{f}) + \frac{\partial}{\partial x_k}(\rho \widetilde{u_k}\tilde{f}) = \frac{\partial}{\partial x_j}\left(\frac{\mu_t}{\sigma_t}\frac{\partial \tilde{f}}{\partial x_k}\right) \qquad (7)$$

$$\frac{\partial}{\partial t}(\rho \tilde{f}''^2) + \frac{\partial}{\partial x_k}(\rho \widetilde{u_k}\tilde{f}''^2) = \frac{\partial}{\partial x_j}\left(\frac{\mu_t}{\sigma_t}\frac{\partial \tilde{f}''^2}{\partial x_k}\right) + C_g \mu_t \left(\frac{\partial \tilde{f}''^2}{\partial x_k}\right)^2 - C_d \rho \frac{\epsilon}{k}\tilde{f}''^2 \qquad (8)$$

Where $\sigma_t$, $C_g$ and $C_d$ are 0.85, 2.86 and 2.0 respectively. More details regarding flamelet modelling can be found in the literature (ANSYS, 2017; Reddy et al., 2015; Saini & De, 2017; Saini et al., 2018).

*The Real Gas Model*

The semi-cryogenic engine operates at supercritical pressure, which has a severe impact on the thermodynamic conditions of the reacting mixture. A lot of equations of state are available in the literature of various complexity for modelling real gas behaviour (Ficarella & De, 2009). But there is a necessity to understand which EoS could better predict the combustion and mixing and overall combustion process.

The general form of pressure P for the cubic EoS models is written as:

$$P = \frac{RT}{V - b + c} - \frac{\alpha}{V^2 + \delta V + \varepsilon} \tag{9}$$

Where

P = absolute pressure (Pa)

V = specific molar volume (m³/kmol)

T = temperature (K)

R = universal gas constant

The coefficients $\alpha$, $b$, $c$, $\delta$, and $\varepsilon$ are given for each EoS as functions of the critical temperature $T_c$, critical pressure $P_c$, acentric factor $\omega$ and critical specific volume $V_c$. The attractive coefficient has a temperature dependence and is commonly written as $\alpha=\alpha(T)$. Detailed descriptions pertaining to individual models can be found in Appendix-A at the end.

*Chemical Kinetics Description Model*

To describe the accurate combustion phenomena in a semi-cryogenic engine, the selection of a reduced chemical kinetics model over the complete plays an important role. In the present work, we have used two chemical kinetics mechanisms. The following section provides a detailed description of the two chemical kinetics model.

(1) **Reduced Jones Lindstedt (JL-R) Mechanism**

Jones Lindstedt (JL) mechanism is a four-step reaction mechanism that is used to describe the combustion phenomena in methane/air mixtures (Jones & Lindstedt, 1988). Later, some modifications were done by Andersen et al. (2009) in order to use the JL mechanism for the modelling of oxy-fuel combustion phenomena. They modified the $H_2 - CO - CO_2$ while retaining initial initiation reactions. This modification helped in better prediction of the concentration of the major species. In addition to this later, some further modifications were done by Frassoldati et al. (2009). Dissociation reactions of $H_2O$ and $O_2$ were introduced in the kinetic mechanism for better prediction of oxy-methane combustion phenomena. A detailed formulation of the JL-R mechanism (De Giorgi et al., 2012) has been reported in Table 2.

(2) **The Detailed Skeletal Mechanism**

The Skeletal (SKEL) model (De Giorgi et al., 2011a, 2011b, 2012) is a reduced form of GRI-Mech 3.0. This mechanism is most suitable for methane-air combustion. To adapt the mechanism for oxy-combustion, it was necessary to eliminate the Nitrogen compounds from the species that were available in this mechanism. Thus, there are 15 species and 41 reactions that are involved in this mechanism. A detailed formulation of the SKEL mechanism (De Giorgi et al., 2012) has been reported in Table 3.

**Geometrical Details, Solver details, and Test condition**

In the present study, the Mascotte test bench RCM03(version 04) setup has been chosen to study the $GCH_4$/LOx combustion characteristics. The G2 test case is one of the most suitable test conditions for such a study. The test condition is summarized in Table 4. The injector consists of an inner and outer diameter of 5 mm and 5.4 mm (diverging duct) for oxygen (oxidizer), respectively. Methane is injected coaxially with an annular duct with an inner and outer diameter of 5.6 mm and 10 mm, respectively. An axisymmetric simulation is performed to maintain a low computational cost. Thus, the chamber section is modeled as a cylinder with a radius of 28.2 mm, which also preserves the chamber section area. The combustion modeling is done at a chamber pressure of 5.6 MPa, which is higher than the critical pressure of oxygen and methane (5.043 and 1.313 MPa, respectively). Methane and liquid oxygen enter at the mass flow rate of 0.0444 kg/s and 0.1431 kg/s, where the density of oxygen corresponds to 1177.8 kg/m3. Methane and liquid oxygen are injected at a temperature of 288

K and 85 K, respectively. The computational domain is shown in figure 1, which extends to 320 mm in the axial direction and 28.2 mm in the radial direction.

ANSYS FLUENT is used to carry out the numerical simulations while the grid generation is carried out in ICEM-CFD. A coupled algorithm is used to solve the momentum and pressure-based continuity equation together that updates the pressure and velocity fields. The convection and diffusion terms are discretized using the second-order upwind scheme and central difference scheme, respectively. The reacting flow inside the combustion chamber is developed at constant chamber pressure by giving a fixed mass flow-rates at fuel inlet and oxidizer inlet. The adiabatic and no-slip condition is imposed on the chamber wall and the near-injector wall. At the domain outlet, the pressure-outlet boundary condition is used with the pressure value as chamber pressure. Figure 1 shows the corresponding boundary conditions.

**Results and discussion**

In this section, we present simulation results at transcritical conditions. We first examine the real gas models, and thereafter we find a better model combination in three different regions, i.e. suitable grid, turbulence model, and chemical kinetics model, which can describe the combustion phenomena accurately at G2 test conditions. The sensitivity of the predictions towards the choice of real gas model, turbulence model, and chemical kinetics is studied in detail. Finally, we analyze the effect of the elevated pressure on the turbulent flame structure of $GCH_4/LOx$ by incorporating a more suitable reduced Jones-Lindstedt oxy-methane chemical kinetics mechanism at elevated pressure.

*Real Gas Model analysis*

Initially, we assess the impact of real gas models to analyze which real gas model can better predict the thermodynamic properties at supercritical pressure. The analysis is validated against the reference data from the NIST webbook (NIST). The computed results in Figure 2 show a very sharp peak of constant pressure specific heat in the vicinity of 195 K (for Methane) and 159 K (for Oxygen). These temperatures correspond to the pseudo-critical temperature for these gases. A sharp decline in density is also observed in this trans-critical regime, as evidenced by Figure 2. The result implies that energy transfer from surrounding hot gas to cryogenic oxygen core through the shear layer would be stored without a noticeable increase in temperature and cause a rapid volume expansion. When compared to the NIST data, we observe better conformity with the SRK EoS results than the other real gas models in the entire temperature range. Hence, the SRK EoS is used in further simulations. Figure 3 depicts that with the increase in the chamber pressure, the slope of the density curve decreases. The decrease in the slope indicates that the pseudo-boiling effect weakens with the increase in the chamber pressure.

*Grid Independence and Error analysis*

To investigate the grid independence at G2 condition as given in Table 4, a simulation is done with a fixed geometry on three different grids: one consists of 33,236 cells; the other two finer grids contain 125,546 and 479,880 cells, respectively. The convergence and stability of the simulations are significantly dependent on the distribution of the grid cells: the concentration of cells should be sufficiently high in the region with steep gradients of physical quantities of the flow. In high-gradient flow regions, the grid should be fine enough to minimize the change of values of flow variables across the cells. Initially, we have carried out some iterations using the fine mesh to find out the optimum value of turbulence model parameters, which is found to be $C_{\varepsilon 1} = 1.4$. During grid independence analysis, we invoke the standard k-ε (SKE) turbulence model with modified dissipation constant ($C_{\varepsilon 1} = 1.4$) with Jones Lindstedt kinetic mechanism and SRK real gas model in conjunction with steady flamelet combustion model.

For the evaluation of the grid, we compare the flame zone against the experiments (De Giorgi et al., 2012; Singla et al., 2005) for each mesh along with the axial velocity profile comparison. Figure 4 reports the predicted results obtained from three different grids for the temperature profile and the axial velocity profile, while Table 5 presents the results from this analysis. The simulations show that the temperature and axial velocity profiles for grid B and grid C are the same regardless of the mesh resolution. Hence, the grid with 125,546 cells (Grid B) is chosen for further simulations.

For further analysis, we take Grid-B as the base grid and approximate the error in Grid-C (fine-grid) compared to Grid-B from Richardson error estimator, defined by,

$$E_1^{Fine} = \frac{\epsilon}{1 - r^o} \tag{10}$$

Error in Grid-A (coarse grid), compared to the solution of Grid-B, is approximated by coarse-grid Richardson error estimator, which is defined as,

$$E_2^{Coarse} = \frac{r^o \epsilon}{1 - r^o} \tag{11}$$

Where the error calculation two consecutive grids are based on the following expression as,

$$\epsilon = \frac{(f_2 - f_1)}{f_1} \tag{12}$$

Grid-Convergence Index (GCI), is calculated, which provides a uniform measure and also accounts for the uncertainty in the Richardson error estimator. GCI for fine-grid and coarse-grid is given as (Roache, 1994, 1997),

$$GCI_{Fine} = F_S \left| E_1^{Fine} \right| \tag{13}$$

$$GCI_{Coarse} = F_S \left| E_2^{Coarse} \right| \tag{14}$$

GCI is calculated for the grid refinement and coarsening of the base grid (Grid-B). A second order accuracy (o=2) with the factor of safety of 1.25 is considered for the best estimation of the grid convergence relating to a 50% grid refinement (coarsening) (Ali et al., 2009). The temperature along the axis is evaluated for grid convergence study and tabulated in Table 6. Error estimation by Richardson extrapolation and GCI for the fine grid is relatively low compared to the coarse grid. Therefore, we choose Grid-B for the rest of the detailed analysis reported herein.

*Turbulence model analysis*

Next, we investigate the effect of different turbulence models, i.e. Standard k-epsilon model (SKE) with modified dissipation constant ($C_{\varepsilon 1}$ = 1.4), Realizable k-epsilon (RKE) model, Reynold stress model (RSM) and k-ω shear stress transport model (k-ω SST) in combination with the flamelet model and JL-R chemical mechanism. This analysis aims to select the best turbulence model, which is able to predict the axial location of the temperature peak within the experimental zone and flame structure at G2 condition.

Figure 5 reports the turbulence model analysis through centreline temperature and density plots. Also, we have plotted the solutions obtained using the ideal-gas assumption. Noticeably, the ideal-gal assumption behaves differently irrespective of the turbulence models, and this behaviour is consistent with our observation reported in the real-gas analysis (Figure 2). As observed, the density magnitude is much lower in the case of ideal EoS; hence, the momentum is also higher, and that diffuses the temperature peak towards downstream locations. This is true for all the turbulence models considered herein. Another point to be noted here is that real EoS is able to capture the pseudo boiling effect on the flame structure which accurately captures the expansion in the radial direction as depicted in temperature profiles. On the other hand, ideal EoS exhibits much longer flame length for all the cases. Secondly, the predictions reveal that that the SKE with a modified constant is able to accurately capture the axial location of temperature peak while comparing to other turbulence models. As observed, the temperature peak has been axially shifted due to higher diffusive effects in all other turbulence models. The reason behind the shifting in centreline temperature peak is also visible form the OH contour plots as depicted in Figure 6. Only, SKE model exhibits consistent OH profiles while comparing with measurements. The rest of the turbulent model show diffused characteristics, and that is the primary reason behind the shifting of centreline peak temperature. Table 7 reports the predicted temperature peak along with flame locations, and one can see that only SKE with the modified constant model is able to accurately capture the flame location. Also, the predicted temperature peak location by SKE with the modified constant model is consistent with the reported literature at this pressure. Hence, SKE model predictions are in good agreement with measurements and show realistic flame characteristics.

Moving forward, we investigate the temperature contours in the whole domain as illustrated in Figure 7. It shows distributions of streamlines too. The figure also shows the transport processes that take place between the jet core and surrounding zone through turbulent transfer. Close observation of the plots reveals the existence of peak contours between cold $O_2$ and flame zone. This happens due to the pseudo boiling phenomenon that causes rapid volumetric expansion in the radial directions at this supercritical pressure. Also, the streamlines plot indicates the existence of a recirculation bubble, which allows to bring the fresh $O_2$ towards the core region. As expected, the temperature contours are quite different for different turbulence models which is consistent with OH predictions shown earlier (Figure 6). These discrepancies arise due to inaccurate estimation of turbulent kinetic energy and dissipation while handling such high-pressure reacting conditions, where the compressibility corrections in turbulence models become extremely critical.

*Chemical Kinetics Mechanism analysis*

In literature, a large number of kinetic schemes for the combustion modelling of methane/air mixture and oxygen-methane are available. Real combustion phenomena involve a very high number of species and reactions. To reduce the computational time, it is important to find a reduced model to be able to reproduce the real phenomena but less heavy for the calculation. The objective of this section is to find a better compromise between the two tendencies using a reduced model compared with a complete one. Thus, we have performed a comparison between the two kinetic mechanisms namely JL-R and SKEL.

Figure 8 illustrates the flame structure analysis for these two mechanisms at stoichiometric scalar dissipation rate ($\chi_{st}$) equal to zero (laminar flamelet calculation). It compares the temperature and mole fraction profiles of different species along the mean mixture fraction ($f$) for the two mechanisms for the flamelet solution. As observed, there is hardly any visible differences in the flamelet solutions of these mechanisms as they produce similar flame structures. Also, the temperature profile, along with the maximum temperature value for the JL-R mechanism, is in agreement with the SKEL mechanism.

While looking at the turbulent calculations in Figure 9, there seems to have some minor differences while flame-turbulence interaction is invoked. However, both the mechanism produces a similar temperature peak and axial velocity along with the inner shear layer. Some minor differences in velocity profiles occur due to the differences in density magnitude in the SKEL mechanism, which is lower compared to JL-R predictions. Further, the flame length and maximum temperature are tabulated in Table 8. Close monitoring of species mass fraction profiles in Figure 10 clearly illustrates that there are no significant differences in the predictions of these two mechanisms, while JL-R is more computationally cost effective. However, one must note that SKEL mechanisms predict similar OH contours as reported in measurements (Figure 11). Though no major differences exist in global parameters, still there exist some subtle differences in minor species predictions like OH. This reveals that the SKEL mechanism accurately predicts the pseudo boiling phenomena that occur at these pressures.

*High Pressure analysis*

Further, we analyse how chamber pressure affects transcritical flame structures while keeping the inlet temperature of the propellant the same as that of the G2 case. The numerical results show that elevated pressure leads to a decrease in flame temperature, as shown in Figure 12. The density gradient and the velocity magnitude are considerably sensitive to the chamber pressure in the narrow oxygen-rich mixing region. By increasing the pressure, which is more than the critical pressure of oxygen (5.043 MPa), the pseudo boiling behaviour becomes less effective. Figure 13 shows the effect of the elevated chamber pressure on the transcritical turbulent flame structure. It is observed that at near critical pressure (5.6 MPa) due to the dominant pseudo-boiling effect, the turbulent flame is much shorter as a result of enhanced turbulent mixing and intense combustion process (Kim et al., 2013). Elevating the chamber pressure from the critical pressure of oxygen (5.043 MPa) weakens the pseudo boiling characteristics, which increase the flame length primarily due to weaker expansion and narrower flame spreading. At the chamber pressure of 8 MPa, the strength of expansion is so weakened, the turbulent mixing process in the shear layer becomes much weaker, and the flame field becomes substantially longer, which tends to increase the flame length. Figure 14 presents the distribution of OH mass fraction at different chamber pressures (5.6 MPa and 8 MPa), signifying the effect of the elevated chamber pressure on the transcritical turbulent flame length. We can observe that the turbulent flame at near critical pressure (5.6 MPa) is much shorter than the flame at the elevated pressure (8 MPa). It can be attributed due to the dominant pseudo-boiling effect at 5.6 MPa. Elevating the chamber pressure from the critical pressure of oxygen (5.043 MPa) weakens the pseudo boiling characteristics, which increase the flame length as observed for pressure 8 MPa. Further, the flame length and maximum temperature at different pressures are tabulated in Table 9.

*LOx Inlet Temperature analysis at 5.6 MPa*

Furthermore, extensive analysis has been carried out to study the effect of the LOx inlet temperature on the transcritical flame structures while keeping the chamber pressure (5.6 MPa) the same as that of the G2 case. The numerical results yield that with the increase in the LOx inlet temperature leads to a decrease in flame temperature for the transcritical LOx inlet injection, as shown in Figure 15. On the contrary, there is an increase in the overall flame temperature when the LOx temperature injection temperature supercritical, i.e. more than the critical temperature of the oxygen. The density gradient is sensitive to the chamber pressure in the narrow oxygen-rich mixing region. By increasing the LOx inlet temperature, the pseudo boiling behavior becomes less effective. The overall volumetric expansion decreases with an increase in the inlet temperature. Figure 16 reports the effect of the elevated inlet temperature on the transcritical turbulent flame structure. It is observed that when the LOx inlet temperature is much less than the critical temperature value of oxygen, the pseudo-boiling effect is more dominant due to which the turbulent flame is much shorter as a result of enhanced turbulent mixing and intense combustion process (Kim et al., 2013). Elevating the inlet temperature decreases the overall volumetric expansion leading to an increase in the entrainment length of the oxygen core. Thus, this phenomenon leads to an increase in the overall length of the flame. Further, the flame length and maximum temperature at different LOx inlet temperatures are tabulated in Table 10.

**Conclusion**

The present work aims to investigate the characteristics of the transcritical turbulent flame structure at the G2 test condition. We aim to find a better model combination in three different regions, i.e. real gas model, turbulence model and chemical kinetics model in combination with the steady β-PDF flamelet model which is able to describe the combustion phenomena accurately at supercritical pressures.

Firstly, we have performed a real gas model analysis. The numerical results indicate that the choice of the real gas model is a critical parameter in simulating the combustion process numerically. Thus, a preliminary analysis indicates that the SRK EoS shows better conformity with the NIST data. Under supercritical pressures, the cryogenic oxygen is dominantly affected by the so-called pseudo boiling behaviour where a steep variation of thermodynamic properties occurs. Further, the turbulence model analysis illustrates that the SKE turbulence model with modified constant ($C_{\varepsilon 1} = 1.4$) is able to accurately predict the flame characteristics (flame location and peak temperature) compared to other turbulence models. Later on, with the chosen turbulence model and EoS, we have carried out chemical kinetic analysis on the flame structure. The simulated results indicate that both the JL-R and SKEL mechanism accurately capture the temperature peak and axial position while exhibiting some differences in OH mass fractions. On the other hand, JL-R is computationally cost effective. Finally, the predictions at elevated chamber pressure and LOx inlet temperature show weakened pseudo boiling behaviour while the flame becomes longer due to narrower flame spreading. This is associated with lower turbulent rates of energy and mass transfer through the transcritical mixing layer.

**Acknowledgment**

Financial support for this research is provided through IITK-Space Technology Cell (STC). The authors would like to acknowledge the IITK computer center ([www.iitk.ac.in/cc](www.iitk.ac.in/cc)) for providing the resources to perform the computation work, data analysis, and article preparation.

**Compliance with Ethical Standards**

We wish to confirm that there are no known conflicts of interest associated with this publication.

## Appendix-A

### Redlich-Kwong (RK) Equation:

The parameter δ is set equal to b, while c and ε are set to 0 in Eq. (9). The function α(T) is given by Eq. (A1). The Redlich-Kwong equation is the simplest of the cubic equations of state and requires two parameters only, the critical temperature Tc and the critical pressure Pc.

$$a(T) = \frac{\alpha_0}{\left[\frac{T}{T_c}\right]^{0.5}} \tag{A1}$$

$$\alpha_0 = \frac{0.42747 R^2 T_c^2}{P_c} \tag{A2}$$

$$b = \frac{0.08664 R T_C}{P_c} \tag{A3}$$

### Soave-Redlich-Kwong (SRK) Equation:

As in the original Redlich-Kwong equation, the parameter δ is set equal to b, while c and ε are set to 0 in Eq. (9). The function α(T) is given by Eq. (A4), and $\alpha_0$ and b are given by Eqs. (A6-A7).

$$\alpha(T) = \alpha_0 (1 + n(1 - T/T_c)^{0.5})^2 \tag{A4}$$

$$n = 0.48 + 1.574\omega - 0.176\omega^2 \tag{A5}$$

$$\alpha_0 = \frac{0.457247 R^2 T_c^2}{P_c} \tag{A6}$$

$$b = \frac{0.07780 R T_C}{P_c} \tag{A7}$$

The Soave-Redlich-Kwong requires three parameters, the critical temperature Tc, the critical pressure Pc, and the acentric factor ω.

### Peng Robinson (PR) Equation:

In the Peng-Robinson equation, δ is set equal to 2b, ε is equal to –b², and c is set to 0. The function α(T) is given by Eq. (A8) with n provided in Eq. (A9) as follows:

$$\alpha(T) = \alpha_0 (1 + n(1 - T/T_c)^{0.5})^2 \tag{A8}$$

$$n = 0.37464 + 1.54226\omega - 0.26992\omega^2 \quad (A9)$$

$$\alpha_0 = \frac{0.457247 R^2 T_c^2}{P_c} \quad (A10)$$

$$b = \frac{0.07780 R T_c}{P_c} \quad (A11)$$

Similar to the Soave-Redlich-Kwong equation, the Peng-Robinson equation is a three-parameter equation and requires the critical temperature $T_c$, the critical pressure $P_c$, and the acentric factor $\omega$.

**The Concept of Volume Translation**

The SRK and PR EoS are cubic in nature and are suitable for vapor molar volume predictions while they provide inaccurate liquid molar volume predictions over a wide range of pressures. Hence the concept of volume translation is implemented to improve the molar volume predictions by shifting the predicted liquid volume by systematic deviation, c, as observed between the predicted molar volume $(v_{EoS})$ and the corresponding experimental value $(v_{exp})$

$$c = (v_{EoS}) - (v_{exp}) \quad (A12)$$

Where c is known as volume translation parameter. For high temperature and high-pressure volume translation c can be expressed as a linear function of reduced temperature, $T_r$,

$$c = A + B.T_r \quad (A13)$$

$$A, B = f(MW, \omega) = k_0 + k_1 \exp\left(\frac{-1}{k_2 MW \omega}\right) + k_3 \exp\left(\frac{-1}{k_4 MW \omega}\right) + k_5 \exp\left(\frac{-1}{k_6 MW \omega}\right) \quad (A14)$$

Where,

| Constants | VT-SRK EoS | VT-PR EoS |
|---|---|---|
| A (cm³/mol) | | |
| $k_0$ | 0.2300 | -4.1034 |
| $k_1$ | 46.843 | 31.723 |
| $k_2$ | 0.0571 | 0.0531 |
| $k_3$ | 23.161 | 188.68 |
| $k_4$ | 0.0003 | 0.0057 |
| $k_5$ | 267.40 | 20.196 |
| $k_6$ | 0.0053 | 0.0003 |
| B (cm³/mol) |  |  |
| $k_0$ | -0.3471 | -0.3489 |
| $k_1$ | -29.748 | -28.547 |
| $k_2$ | 0.0644 | 0.0687 |
| $k_3$ | -347.04 | -817.73 |
| $k_4$ | 0.0010 | 0.0007 |
| $k_5$ | -88.547 | -65.067 |
| $k_6$ | 0.0048 | 0.0076 |

**Volume Translated-SRK Equations**

The concept of volume translation is implemented in the SRK cubic EoS and can be written as follows

$$P = \frac{RT}{v + c - b} - \frac{\alpha}{(v + c)(v + c + b)} \tag{A15}$$

**Volume Translated-PR Equations**

The concept of volume translation is implemented in the PR cubic EoS and can be written as follows

$$P = \frac{RT}{v + c - b} - \frac{\alpha}{(v + c)(v + c + b) + b(v + c - b)} \tag{A16}$$

where *a* and *b* are dependent on the attractive and repulsive forces amongst molecules and are determined by the critical pressure and critical temperature.

**Table 1.** Different Chemical Kinetic Mechanism.

| Chemical Kinetic mechanism | Number of Species | Number of reactions | Valid For |
|---|---|---|---|
| Reduced JL Mechanism (De Giorgi et al., 2012; Frassoldati et al., 2009) | 9 | 6 | $CH_4$-$O_2$ |
| SKEL Mechanism (De Giorgi et al., 2012; Yang et al., 1998) | 15 | 41 | $CH_4$-$O_2$ |

**Table 2.** Reactions of the modified Jones-Lindstedt kinetics mechanism (Frassoldati et al., 2009).

| Title | Reactions | Reaction Exponent |
|---|---|---|
| JL-R1 | $CH_4 + \frac{1}{2} O_2 \rightarrow CO + 2H_2$ | $[CH_4]^{0.5} [O_2]^{1.3}$ |
| JL-R2 | $CH_4 + H_2O \rightarrow CO + 3H_2$ | $[CH_4][H_2O]$ |
| JL-R3 | $CO + H_2O \Leftrightarrow CO_2 + H_2$ | $[CO][H_2O]$ |
| JL-R4 | $H_2 + \frac{1}{2} O_2 \Leftrightarrow H_2O$ | $[H_2]^{0.3} [O_2]^{1.55}$ |
| JL-R5 | $O_2 \Leftrightarrow 2O$ | $[O_2]$ |
| JL-R6 | $H_2O \rightarrow H + OH$ | $[H_2O]$ |

**Table 3.** Reactions of the skeletal kinetics mechanism.

| Title | Reactions | Title | Reactions |
|---|---|---|---|
| SK-1 | $H + O_2 \rightarrow OH + O$ | SK-22 | $CH_2O + OH \rightarrow HCO + H_2O$ |
| SK-2 | $O + H_2 \rightarrow OH + H$ | SK-23 | $CH_2O + O_2 \rightarrow HCO + HO_2$ |
| SK-3 | $OH + H_2 \rightarrow H_2O + H$ | SK-24 | $CH_2O + CH_3 \rightarrow HCO + CH_4$ |
| SK-4 | $OH + OH \rightarrow O + H_2O$ | SK-25 | $CH_2O + M \rightarrow HCO + H + M$ |
| SK-5 | $H + H + M \rightarrow H_2 + M$ | SK-26 | $CH_3 + O \rightarrow CH_2O + H$ |
| SK-6 | $H + OH + M \rightarrow H_2O + M$ | SK-27 | $CH_3 + OH \rightarrow CH_2O + H_2$ |
| SK-7 | $H + O_2 + M \rightarrow HO_2 + M$ | SK-28 | $CH_3 + O_2 \rightarrow CH_3O + O$ |
| SK-8 | $HO_2 + H \rightarrow OH + OH$ | SK-29 | $CH_3 + O_2 \rightarrow CH_2O + OH$ |
| SK-9 | $HO_2 + H \rightarrow H_2 + O_2$ | SK-30 | $CH_3 + HO_2 \rightarrow CH_3O + OH$ |

| | | |  | | |
|---|---|---|---|---|---|
| SK-10 | $HO_2 + O \rightarrow O_2 + OH$ | | SK-31 | $CH_3 + HCO \rightarrow CH_4 + CO$ |
| SK-11 | $HO_2 + OH \rightarrow H_2O + O_2$ | | SK-32 | $CH_3(+M) + O_2 \rightarrow CH_3 + H(+M)$ |
| SK-12 | $H_2O_2 + M \rightarrow OH + OH + M$ | | SK-33 | $CH_4 + H \rightarrow CH_3 + H_2$ |
| SK-13 | $CO + OH \rightarrow CO_2 + H$ | | SK-34 | $CH_4 + O \rightarrow CH_3 + OH$ |
| SK-14 | $CO + O + M \rightarrow CO_2 + M$ | | SK-35 | $CH_4 + O_2 \rightarrow CH_3 + HO_2$ |
| SK-15 | $HCO + H \rightarrow H_2 + CO$ | | SK-36 | $CH_4 + OH \rightarrow CH_3 + H_2O$ |
| SK-16 | $HCO + O \rightarrow OH + CO$ | | SK-37 | $CH_4 + HO_2 \rightarrow CH_3 + H_2O_2$ |
| SK-17 | $HCO + OH \rightarrow H_2O + CO$ | | SK-38 | $CH_3O + H \rightarrow CH_2O + H_2$ |
| SK-18 | $HCO + O_2 \rightarrow HO_2 + CO$ | | SK-39 | $CH_3O + OH \rightarrow CH_2O + H_2O$ |
| SK-19 | $HCO + M \rightarrow H + CO + M$ | | SK-40 | $CH_3O + O_2 \rightarrow CH_2O + HO_2$ |
| SK-20 | $CH_2O + H \rightarrow HCO + H_2$ | | SK-41 | $CH_3O + M \rightarrow CH_2O + H + M$ |
| SK-21 | $CH_2O + O \rightarrow HCO + OH$ | | | |

**Table 4.** Test conditions for G2 case (Singla et al., 2005).

| | Mass flow rate (kg/sec) | T (K) | ρ (kg/m$^3$) | u (m/s) |
|---|---|---|---|---|
| LOx | 0.0444 | 85 | 1177.8 | 3.70 |
| CH$_4$ | 0.1431 | 288 | 43.344 | 63.2 |

**Table 5.** Predicted values for the Temperature and the Flame location for each grid at 5.6 MPa.

| Grid | Cells | Nodes | Temperature (K) | Flame Location (m) |
|---|---|---|---|---|
| A | 33,236 | 33,681 | 1750.00 | 0.0560 |
| B | 125,546 | 126,425 | 1800.08 | 0.0550 |
| C | 479,880 | 481,613 | 1800.08 | 0.0550 |

**Table 6.** Richardson error estimation and Grid-Convergence Index for three sets of grids.

| | $r_{AB}$ | $r_{BC}$ | o | $\epsilon_{AB}$ | $\epsilon_{BC}$ | $E_{AB}^{Coarse}$ | $E_{BC}^{Fine}$ | $GCI_{Coarse}$ (%) | $GCI_{Fine}$ (%) |
|---|---|---|---|---|---|---|---|---|---|
| **Temperature at $X = 0.0555$** | 2 | 2 | 2 | 0.0510 | 0.0347 | -0.0680 | -0.0115 | 8.5 | 1.44 |

**Table 7.** Predicted values for the Temperature and the Flame location for different turbulence models at 5.6 MPa.

| Turbulence Model | Chemikin Mechanism | Flame Length (m) | Temperature (K) |
|---|---|---|---|
| RKE | JL-R | 0.0792 | 2709.26 |
| RSM | JL-R | 0.0966 | 1492.86 |
| **SKE** | **JL-R** | ***0.0550*** | ***1800.08*** |
| *k-ω* SST | JL-R | 0.0855 | 1580.31 |

**Table 8.** Predicted values for the Temperature and the Flame location for different Chemikin mechanism at 5.6 MPa.

| Turbulence Model | Chemikin Mechanism | Flame Length (m) | Temperature (K) |
|---|---|---|---|
| SKE | JL-R | 0.0550 | 1800.08 |
| SKE | SKEL | 0.0550 | 1895.42 |

**Table 9.** Predicted values for the Temperature and the Flame location at different Pressures.

| Turbulence Model | Chemikin Mechanism | Pressure | Flame Length (m) | Temperature (K) |
|---|---|---|---|---|
| SKE | JL-R | 5.6 MPa | 0.0550 | 1800.08 |
| SKE | JL-R | 8 MPa | 0.064 | 1629.13 |

**Table 10.** Predicted values for the Temperature and the Flame location at different LOx inlet temperature at 5.6 MPa.

| Turbulence Model | Chemikin Mechanism | LOx Inlet Temperature | Flame Length (m) | Temperature (K) |
|---|---|---|---|---|
| SKE | JL-R | 85 K | 0.0550 | 1800.08 |
| SKE | JL-R | 100 K | 0.0580 | 1782.76 |
| SKE | JL-R | 150 K | 0.0630 | 1779.94 |
| SKE | JL-R | 200 K | 0.1134 | 2161.30 |

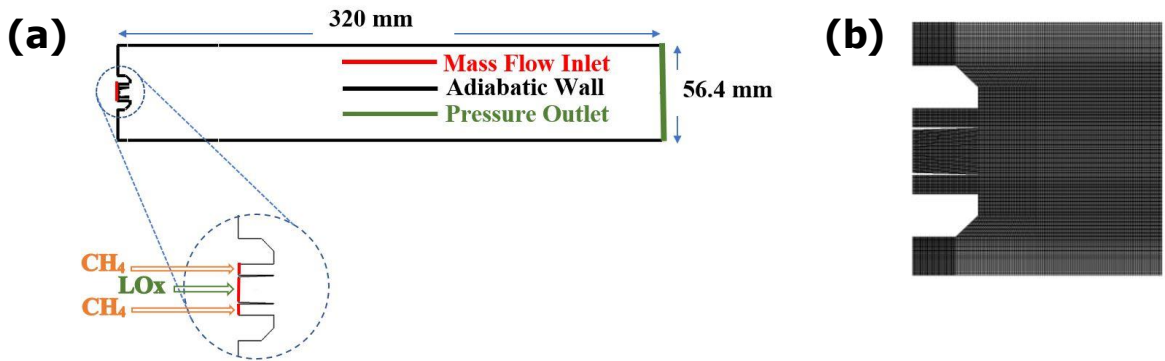

**Figure 1.** (a) Schematic geometry of injector and chamber for the G2 test case (b) Computational Grid for the zoomed section

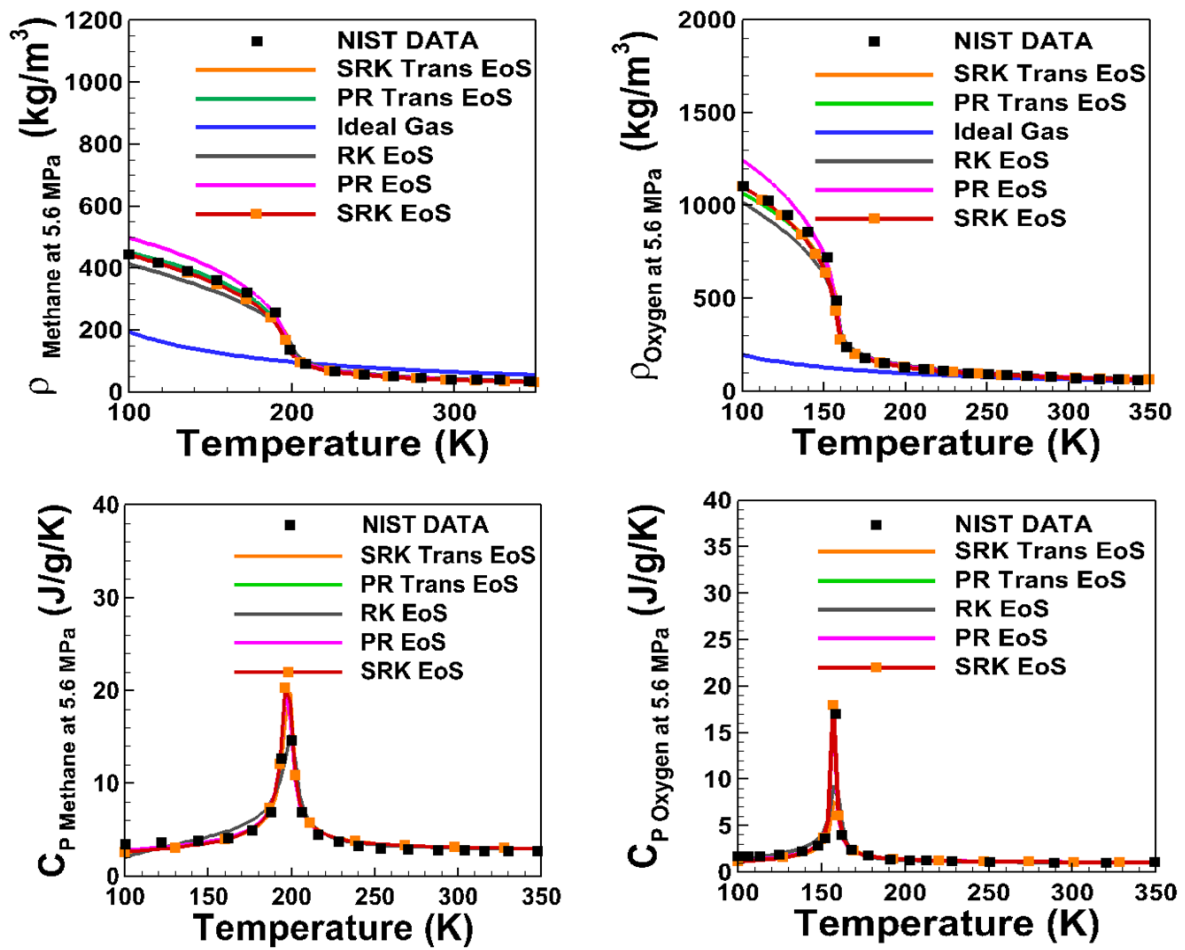

**Figure 2.** Comparison of NIST data and predicted thermodynamic properties for Methane, and Oxygen at 5.6 MPa.

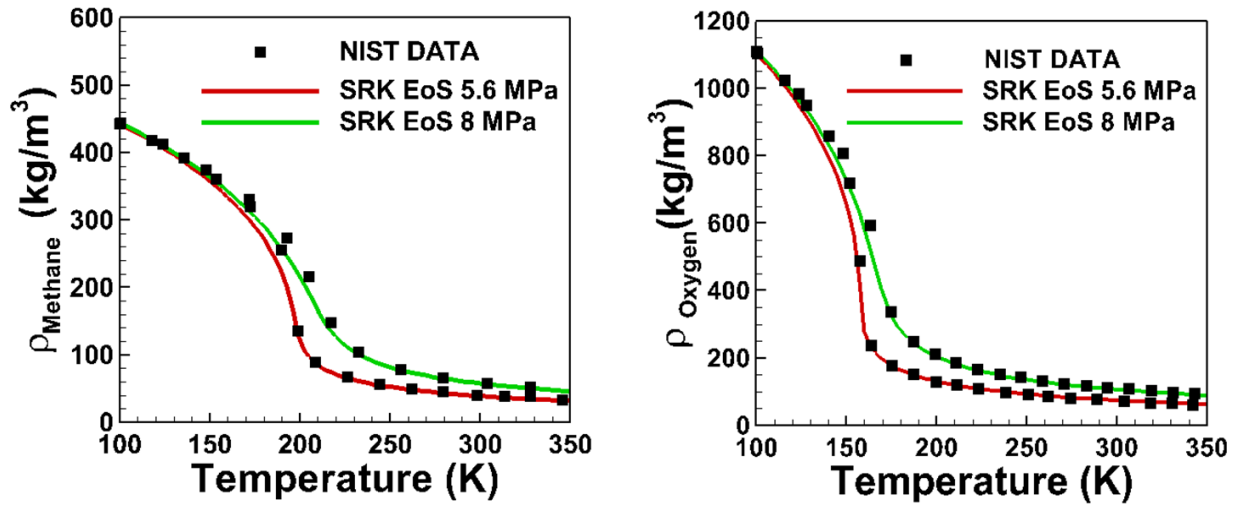

**Figure 3.** Comparison of NIST data and predicted thermodynamic property (density) at 5.6 MPa and 8 MPa for Methane and Oxygen

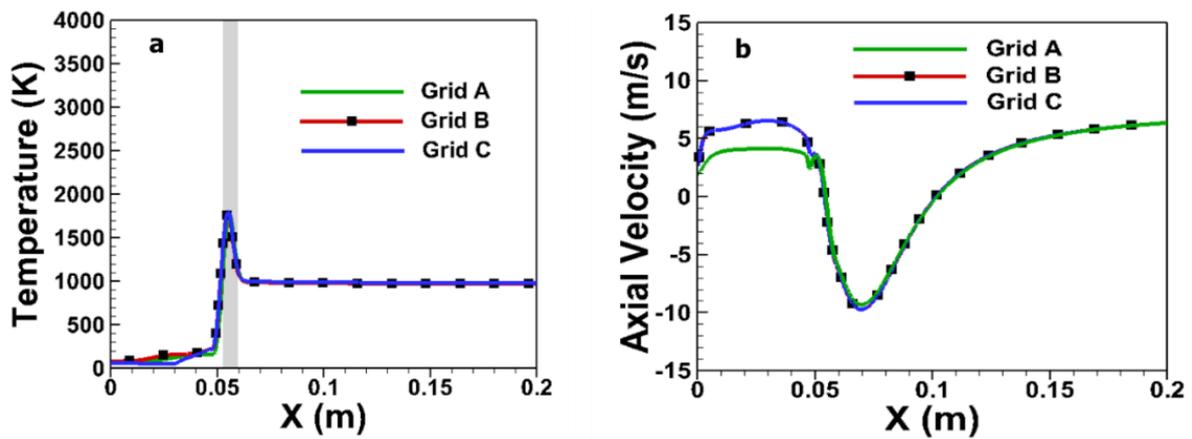

**Figure 4.** Grid Independence study (a) Temperature profile (b) Axial velocity profile

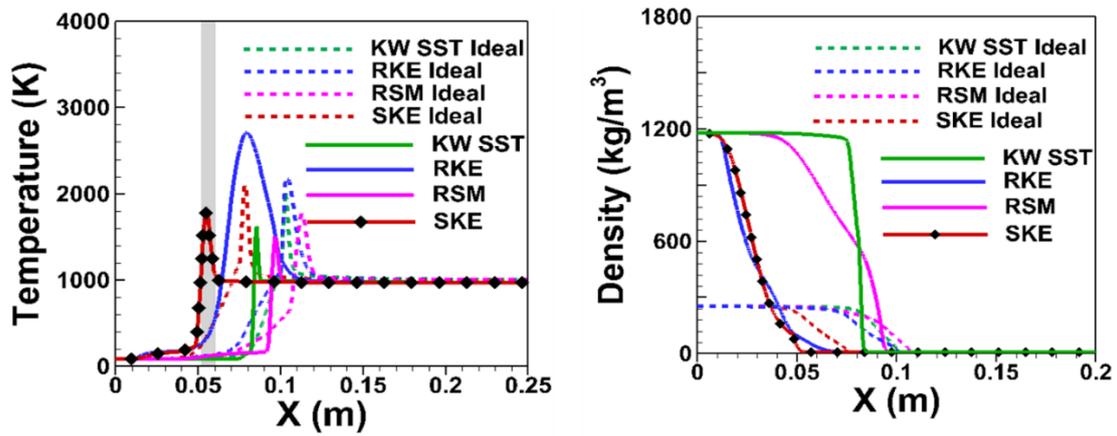

**Figure 5.** Comparison of Temperature and density profiles along the axis of symmetry for different turbulence models

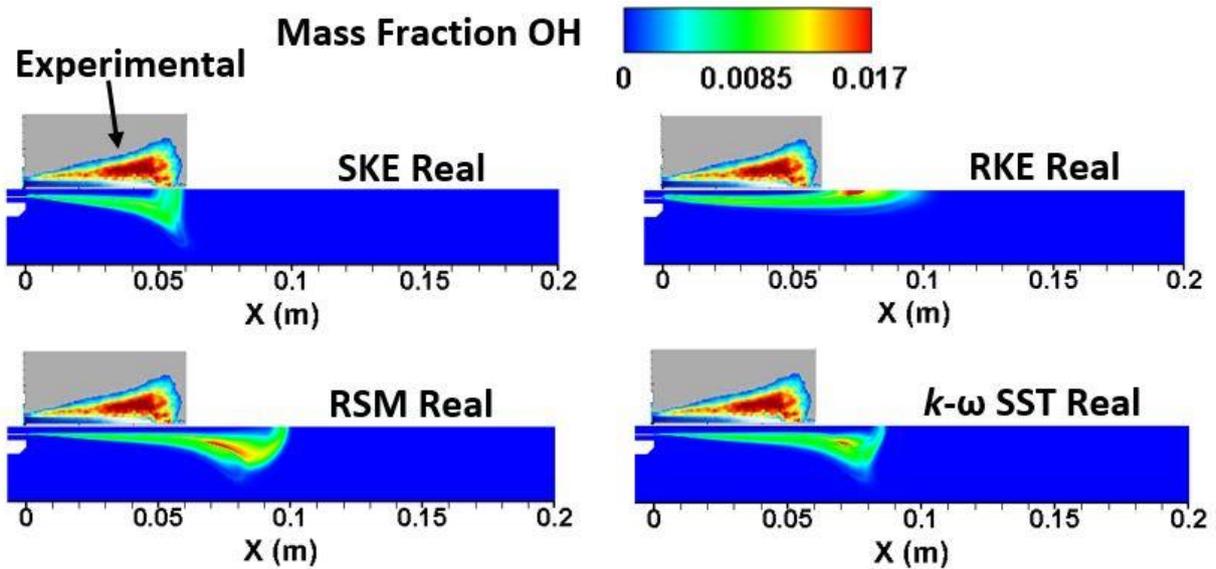

**Figure 6.** Comparison of OH mass fraction contours for different turbulence models at 5.6 MPa

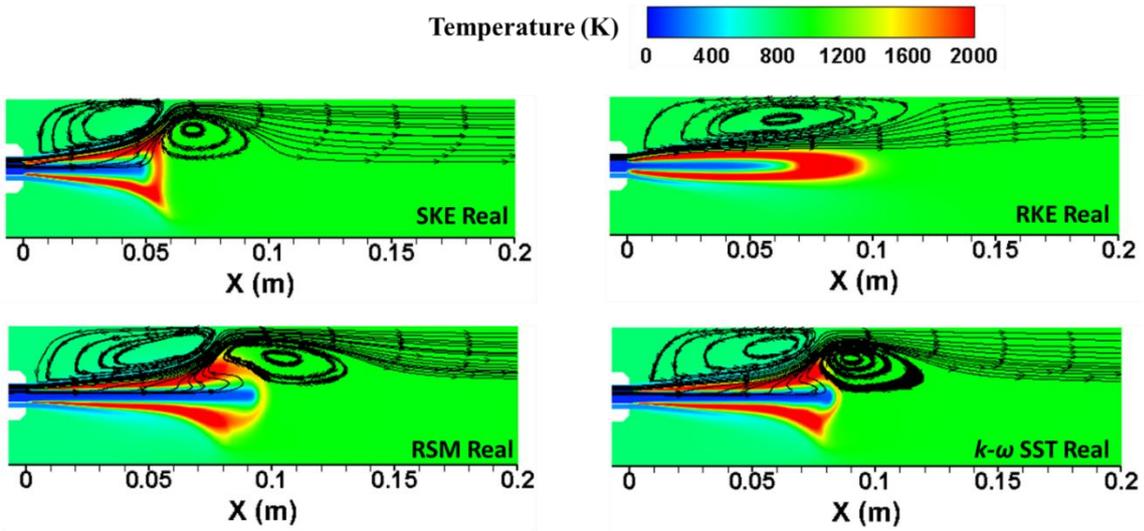

**Figure 7.** Comparison of mean temperature contours for different turbulence models at 5.6 MPa

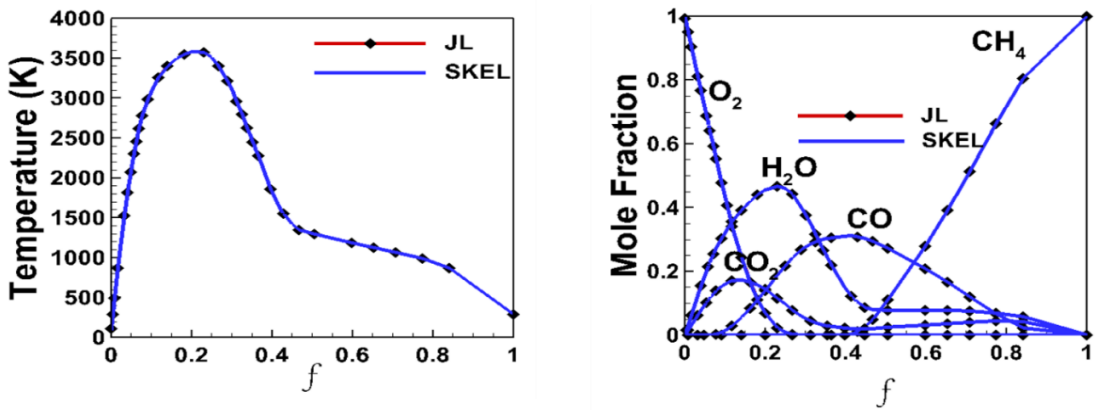

**Figure 8.** Comparison of flame structures for different chemical mechanims (laminar flamelet calculation) at 5.6 MPa

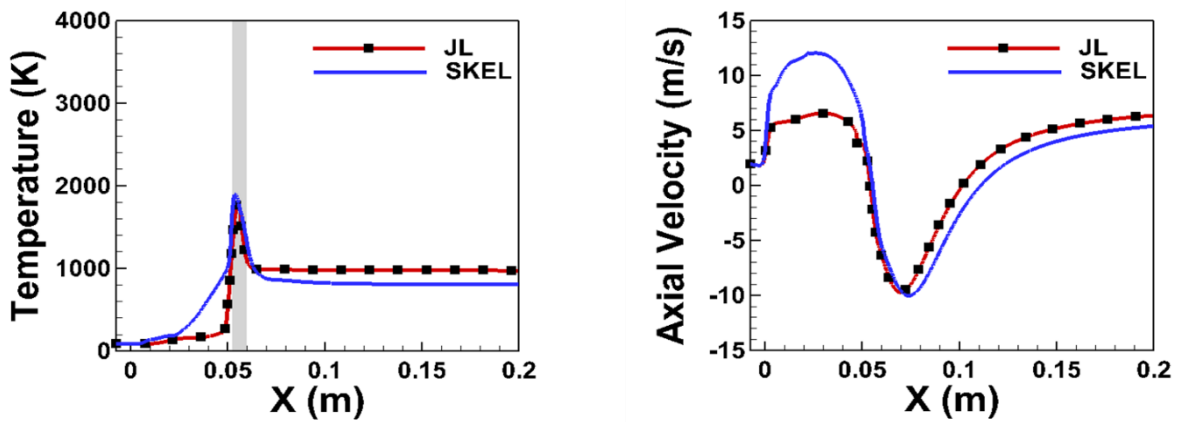

**Figure 9.** Comparison of mean temperature and axial velocity profiles along the centerline

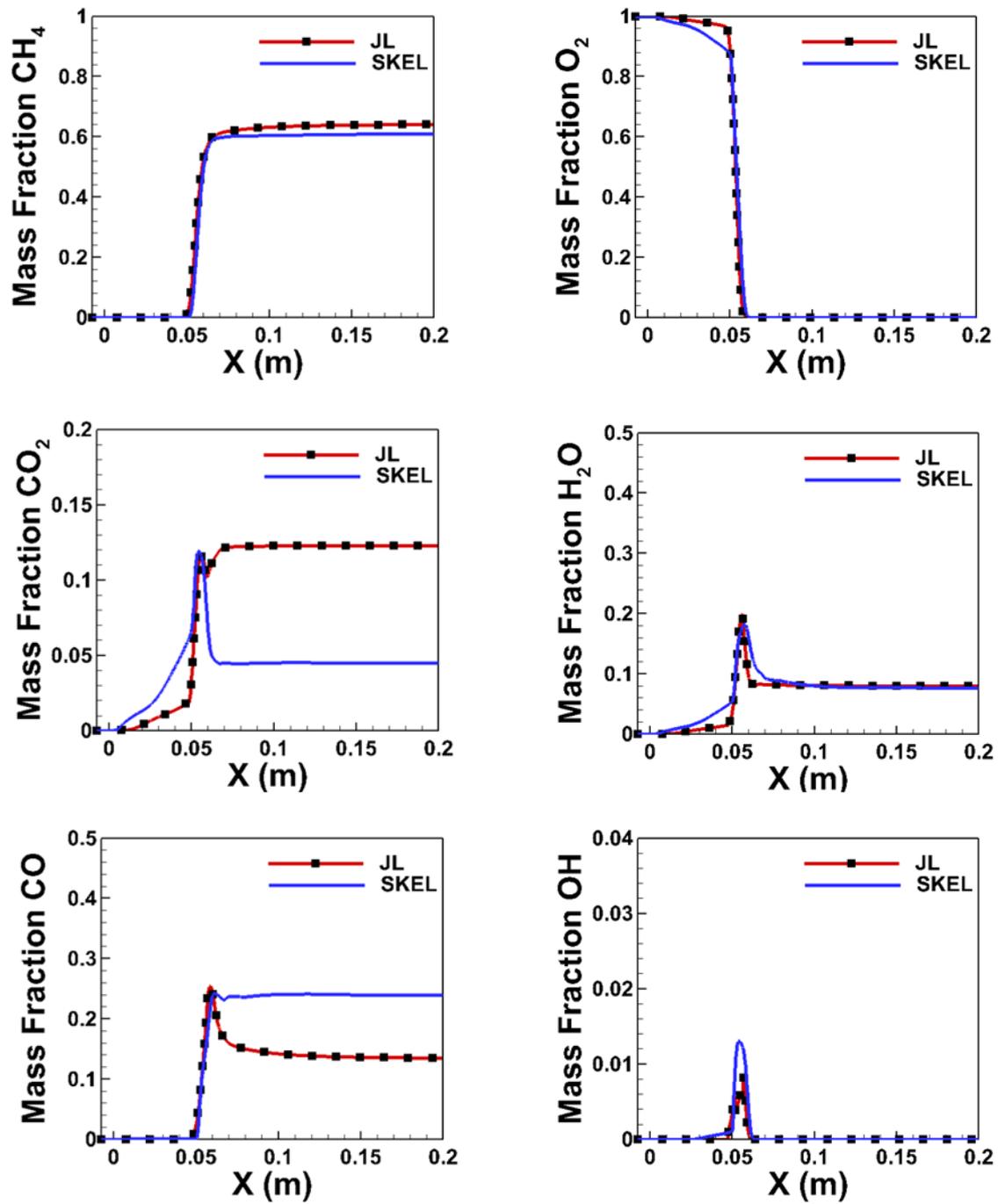

**Figure 10.** Comparison of species mass fractions along the centerline for different chemical mechanisms at 5.6 MPa

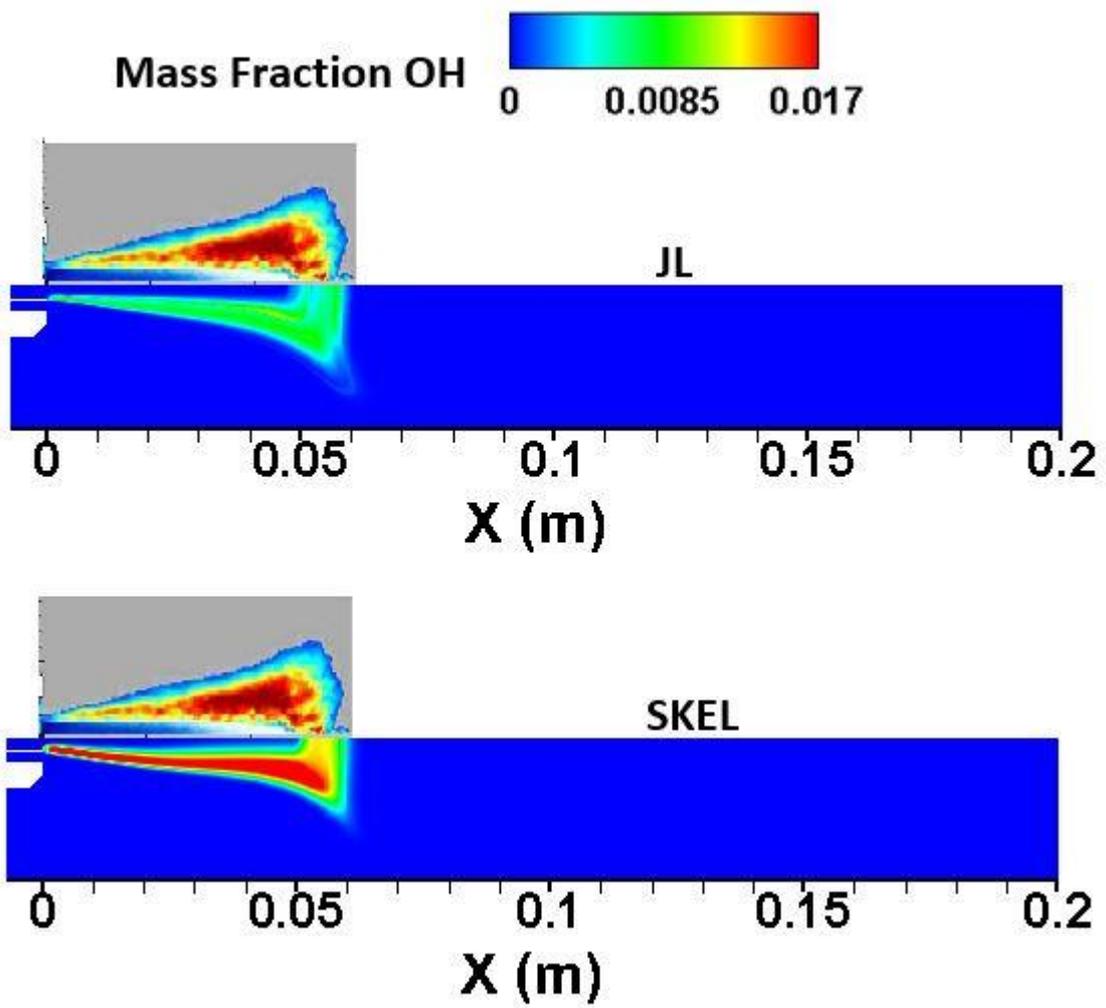

**Figure 11.** Comparison of OH contours for different chemical mechanisms at 5.6 MPa

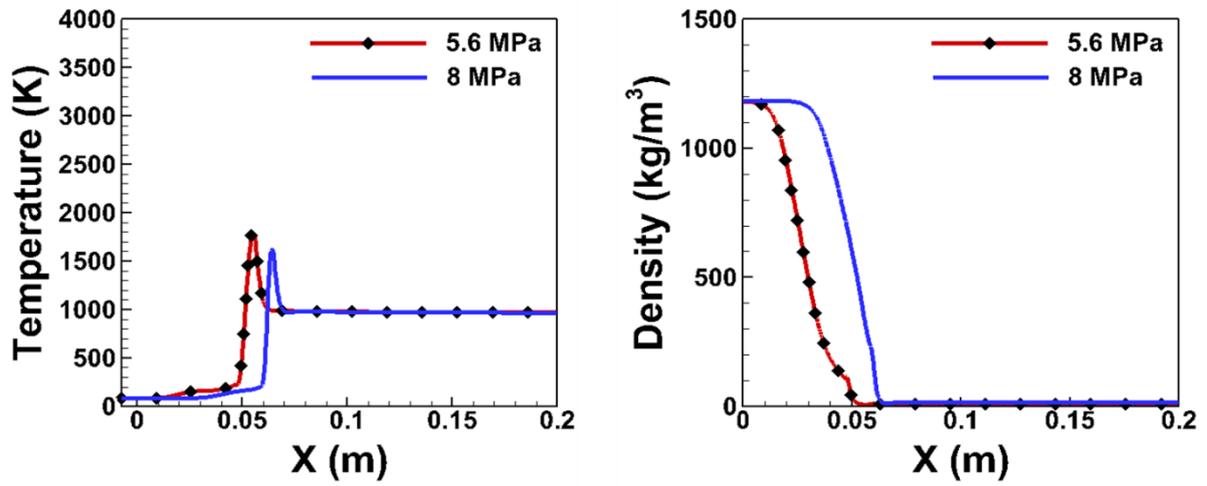

**Figure 12.** Effect of pressure on the Temperature and Density profile

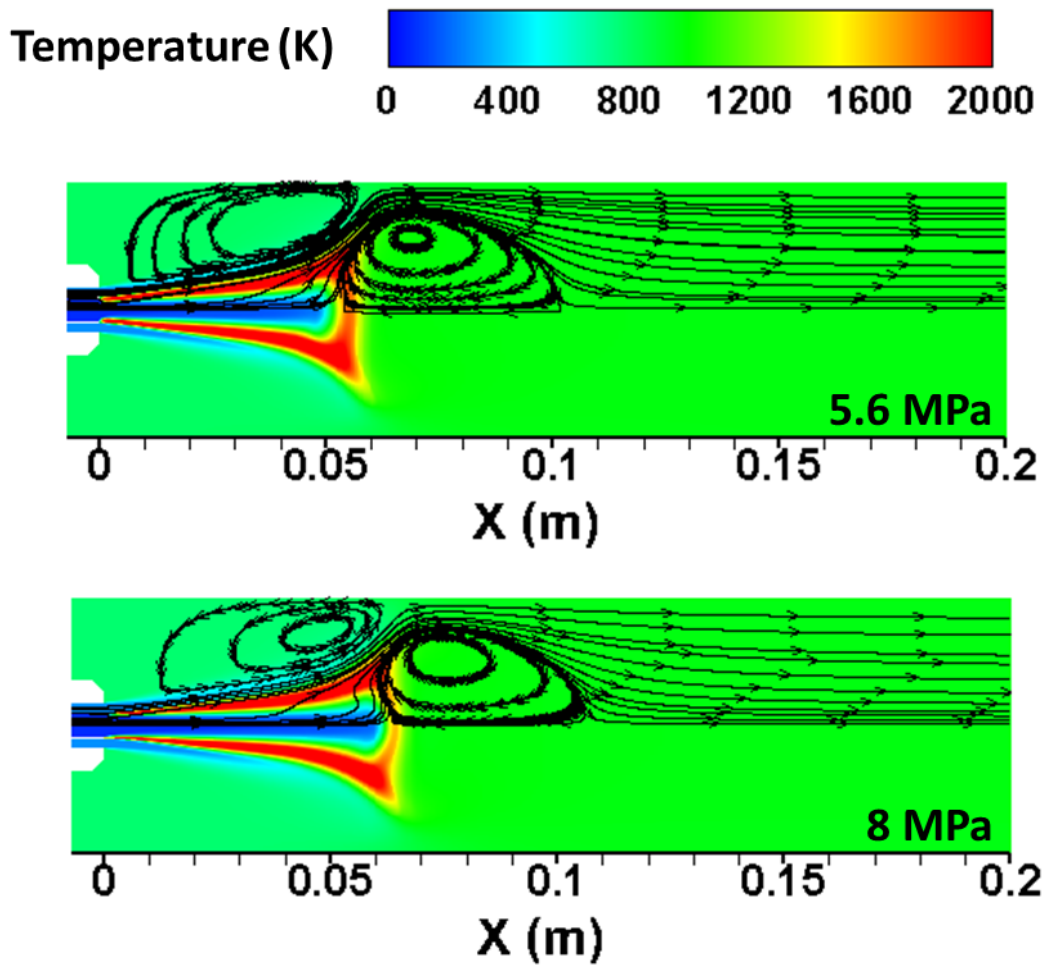

**Figure 13.** Effect of pressure on the temperature at 5.6 MPa, and 8 MPa

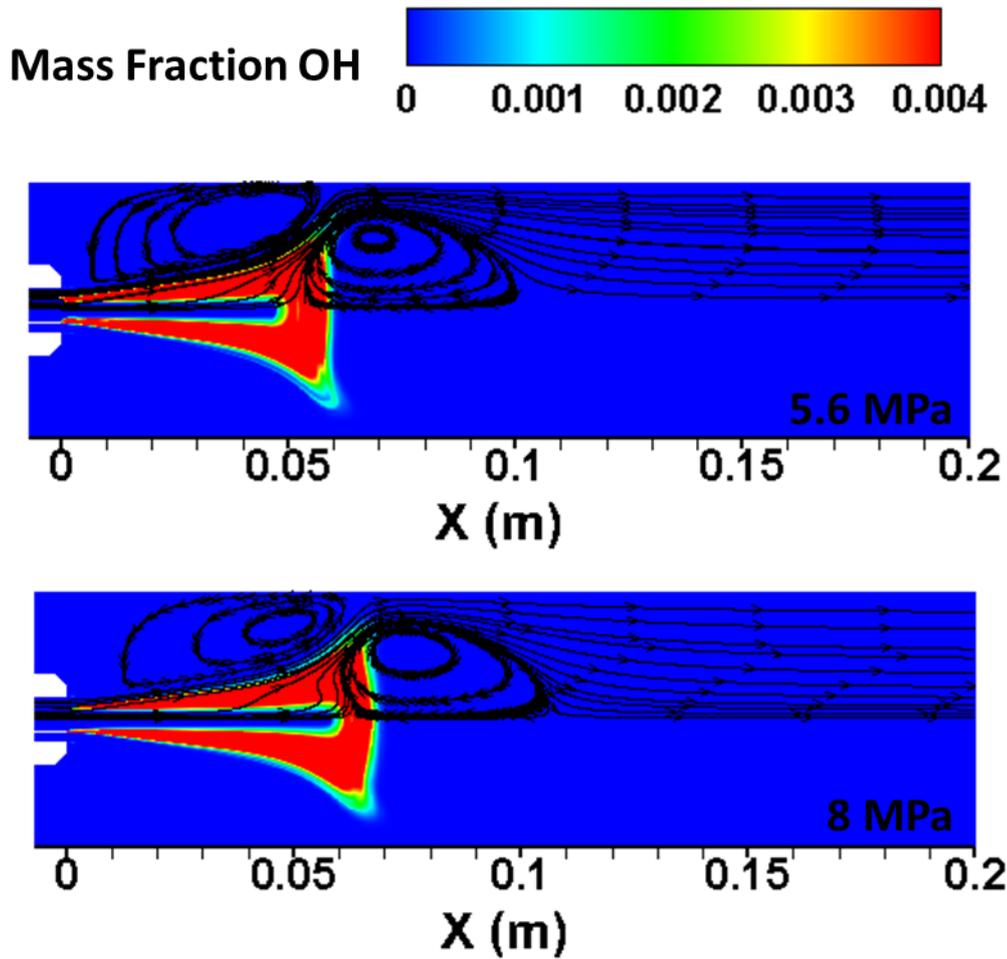

**Figure 14.** Effect of pressure on OH contours at 5.6 MPa, and 8 MPa

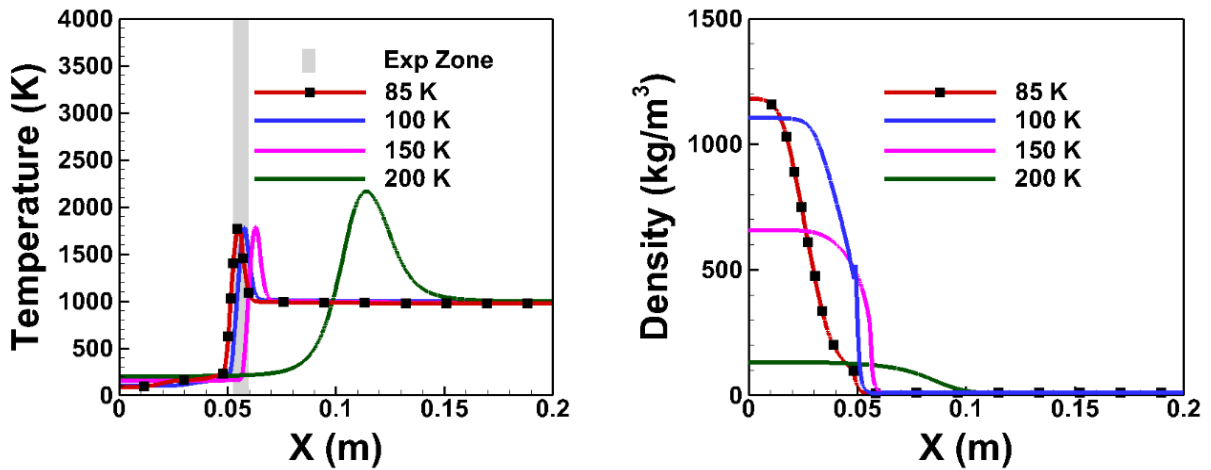

**Figure 15.** Effect of LOx inlet temperature on the Temperature and Density profile at 5.6 MPa

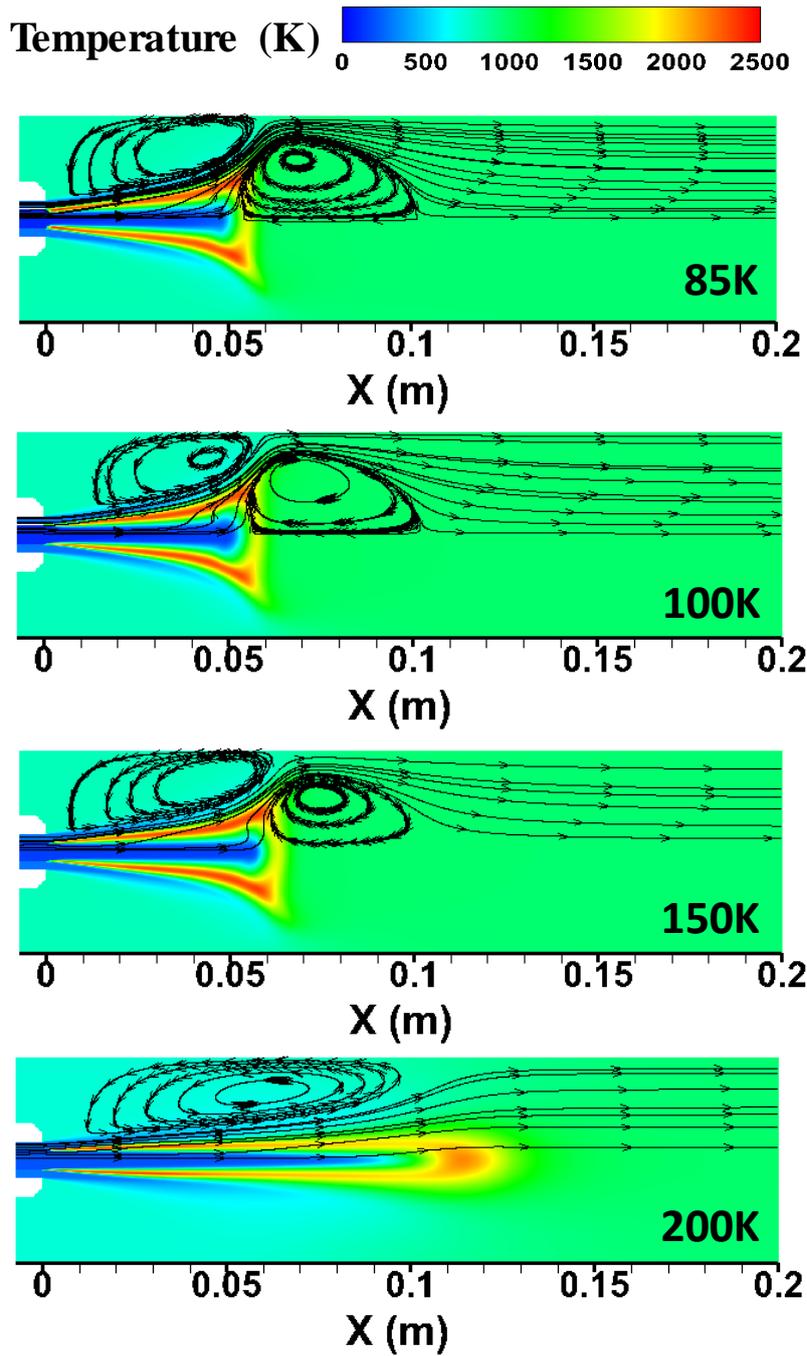

**Figure 16.** Effect of LOx inlet temperature on the temperature contour at 5.6 MPa